\documentclass[aps,prl,twocolumn,showpacs,showkeys,nofootinbib,superscriptaddress]{revtex4-1}

\usepackage[utf8]{inputenc}
\usepackage{graphicx}
\usepackage{amsfonts}
\usepackage{amssymb}
\usepackage{amsmath}
\usepackage[mathscr]{eucal}
\usepackage{setspace}
\usepackage{booktabs}
\usepackage[shellescape,latex]{gmp}
\usempxpackage{amssymb}

\def\bb{\mathbb}
\def\sqr#1#2{{\vcenter{\hrule height.#2pt
   \hbox{\vrule width.#2pt height#1pt \kern#1pt
      \vrule width.#2pt}
   \hrule height.#2pt}}}

\def\bsqr#1#2{{\vrule width #1pt height#2pt}}
\def\bsquare{{\mathchoice\bsqr66\bsqr66\bsqr33\bsqr33}}
\def\badbreak{\penalty1000}

\def\R{{\bb R}}				    
\newcommand{\cP}{{\cal P}}                    
\newcommand{\cS}{{\cal S}}                    
\def\fir{{\scriptscriptstyle{\text{\rm IR}}}}             
\def\fuv{{\scriptscriptstyle{\text{\rm UV}}}}             
\def\smax{{\scriptstyle{\text{\rm max}}}}          
\def\smin{{\scriptstyle{\text{\rm min}}}}            
\def\lm0{{\lambda_0}}                                     
\def\nrN{N}                                               
\def\cf{\mathfrak{n}}                                     
\def\cfu{\cf_\star}                                       
\def\efN{\mathscr{N}}                                     
\def\efNm{\efN_\star}                                     
\def\Obj{O}                                                     
\def\w{c}                                                         
\def\v{b}                                                            

\def\lamA{\lambda_{\scriptscriptstyle{\text{\rm A}}}}        

\newcommand*{\GtrApprox}{\smallrel\gtrapprox}

\makeatletter
\newcommand*{\smallrel}[2][.8]{%
  \mathrel{\mathpalette{\smallrel@{#1}}{#2}}%
}
\newcommand*{\smallrel@}[3]{%
  \sbox0{$#2\vcenter{}$}%
  \dimen@=\ht0 %
  \raise\dimen@\hbox{%
    \scalebox{#1}{%
      \raise-\dimen@\hbox{$#2#3\m@th$}%
    }%
  }%
}
\makeatother

\usepackage{amsmath} 
\usepackage{dsfont}  
\usepackage{bm}      

\def\beq{\begin{equation}}
\def\eeq{\end{equation}}
\def\beqs#1\eeqs{\beq\begin{split} #1 \end{split}\eeq}

\long\def\comment#1{}

\def\be{\begin{equation}}
\def\ee{\end{equation}}

\def\bc{\begin{center}}
\def\ec{\end{center}}

\usepackage{microtype}
\usepackage[dvipsnames]{xcolor}
\usepackage[colorlinks=true,backref=false, linktocpage=true,
citecolor=PineGreen,urlcolor=PineGreen,linkcolor=PineGreen,pdfpagemode=UseOutlines]{hyperref}

\hypersetup{%
  bookmarksnumbered=true,
  pdftitle = {},
  pdfsubject = {},
  pdfauthor = {},
  pdfkeywords = {}
}

\begin{document}

\title{Topological Dimensions from Disorder and Quantum Mechanics?}

\author{Ivan Horv\'{a}th}
\email{ihorv2@g.uky.edu}
\affiliation{Nuclear Physics Institute CAS, 25068 \v{R}e\v{z} (Prague), Czech Republic}
\affiliation{University of Kentucky, Lexington, KY 40506, USA}

\author{Peter Marko\v{s}}
\email{peter.markos@fmph.uniba.sk}
\affiliation{Dept. of Experimental Physics, Faculty of Mathematics, 
Physics and Informatics, Comenius University in Bratislava, Mlynsk\'a Dolina 2, 
842 28 Bratislava, Slovakia}

\date{Dec 18, 2022}

\begin{abstract}

We have recently shown that critical Anderson electron in $D\!=\!3$ dimensions 
effectively occupies a spatial region of infrared (IR) scaling dimension 
$d_\fir \!\approx\! 8/3$. 
Here we inquire about the dimensional substructure involved. We partition space 
into regions of equal quantum occurrence probability, such that points comprising 
a region are of similar relevance, and calculate the IR scaling dimension 
$d$ of each. This allows us to infer the probability density $p(d)$ for dimension 
$d$ to be accessed by~electron. 
We find that $p(d)$ has a strong peak at $d$ very close to 2. In fact, 
our data suggests that $p(d)$ is non-zero on the interval 
$[d_\smin, d_\smax] \!\approx \! [4/3,8/3]$ and may develop a discrete part 
($\delta$-function) at $d\!=\!2$ in infinite-volume~limit. The latter 
invokes the possibility that combination of quantum mechanics and pure disorder 
can lead to emergence of topological dimensions.
Although $d_\fir$ is based on effective counting of which 
$p(d)$ has no a priori knowledge, $d_\fir \!\ge\! d_\smax$ is an exact feature 
of the ensuing formalism. Possible connection of our results to recent findings 
of $d_\fir \!\approx\! 2$ in Dirac near-zero modes of thermal quantum 
chromodynamics is emphasized.
 


\end{abstract}

\maketitle

\noindent

{\bf 1. Introduction. $\,$} 
Understanding spatial geometry~of Anderson transitions~\cite{Anderson:1958a}
is an intriguing problem. Indeed, although studied quite extensively, 
the complicated structure of critical electronic states (see e.g.~\cite{Romer:2008})
leaves room for new insights. Novel characterization may reveal 
unknown details of disorder-driven metal-insulator transitions and, for example, 
lead to deeper understanding of their renormalization group 
description~\cite{Abrahams:1979a}.

Another reason to study the geometry of Anderson~transitions arises by seeing 
them as quantum {\em dimension transitions}, a viewpoint taken in 
Ref.~\cite{Horvath:2021zjk}. Using~effective number theory 
(ENT)~\cite{Horvath:2018aap,Horvath:2018xgx}, which entails a unique 
measure-based dimension 
$d_\fir$~\cite{Alexandru:2021pap, Horvath:2022ewv} for spaces with 
probabilities, it showed that the transition is a two-step dimension reduction
\begin{equation}
    d_\fir \;=\; 3   \;\,\longrightarrow\;\; \, \approx 8/3 \;\,\longrightarrow\;\,  0    
     \label{eq:006}             
\end{equation}
Here the flow is from extended to critical to localized state, and exponential 
localization was assumed. Remarkable property of the above is that these 
reductions are complete~\cite{Horvath:2022lbj}. Indeed, probability doesn't leak 
away from subdimensional effective supports, and electron is fully confined 
to them in infinite volume. It is thus meaningful to say that the space available 
to quantum particle collapses into lower dimensional one under the influence 
of strong enough disorder. As such, it represents a mechanism for generating 
lower-dimensional spaces by simple combination of quantum mechanics 
and disorder.  
 
While dimension is the most basic characteristic of space available to critical 
electron, this space may contain subsets with dimensions $d \!<\! d_\fir$. 
Such substructure may be physically significant if electron mostly resides there. 
The aim of this work is to characterize the critical spatial geometry in such 
manner: we will compute the probability distribution $p(d)$ that electron is 
present in space of dimension $d$. We refer to $p(d)$ as 
{\em dimension content} of Anderson criticality or that of probability 
distribution~in~general.

Critical states at Anderson transitions were recognized to have fractal-like 
features long ago, first interpreting them in analogy to scale-invariant 
fractals~\cite{Aoki_1983, Soukulis:1984a} and later to more complex
multifractals 
\cite{Castellani:1986, Evangelou_1990, Schreiber:1991a, Janssen:1994}. 
Formalism used in the latter mimics one that describes ultraviolet (UV) 
measure singularities occurring in turbulence and strange attractors 
(see e.g.~\cite{falconer2014fractal, Halsey_Kadanoff_multif:1986}).
More recent works in the Anderson context are
\cite{Mildenberger:2002a, Vasquez:2008a, Rodriguez:2008a, Rodriguez:2011, Ujfalusi:2015a}.
However, the focus of multifractal analysis doesn't make it convenient for 
computing $p(d)$. We thus proceed by proposing a method that organizes 
the calculation in terms of probabilities from the outset and zooms in on 
dimensions by degree of their actual presence. Moreover, $d$ involved is 
simply the IR Minkowski dimension of a subset, and thus manifestly 
a measure-based dimension of space. 
In the ensuing multidimensionality formalism, wave function is
\begin{equation}
   \begin{aligned} 
   \text{subdimensional if}&  \quad\quad  d_\fir < D \quad \\
   \text{multidimensional if}&  \quad\quad p(d) \ne \delta(d-d_\smax) \quad \\
   \text{of proper dimension if}&  \quad\quad d_\fir = d_\smax 
   \end{aligned}
   \label{eq:016}   
\end{equation}
where $d_\smax \!=\! \sup\,\{ d \!\mid\! p(d) > 0 \}$, $D \!=\! 3$ is the IR 
dimension of the underlying space, and $d_\fir \ge d_\smax$ holds in general. 

\begin{figure}[t]
   \includegraphics[width=0.40\textwidth]{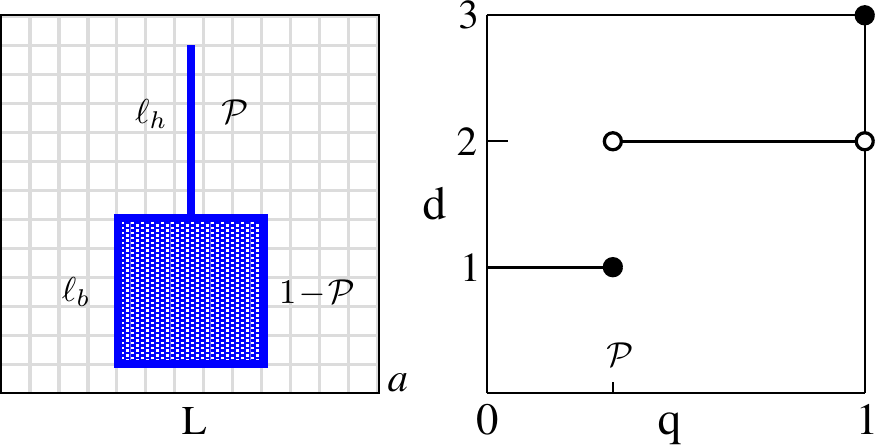}
   \vskip -0.05in   
   \caption{The ``shovel" (left) and $d(q)$ (right) associated with its UV 
   dimension content in $\R^3$. See discussion in the text.}
   \label{fig:illustr1}
   \vskip -0.16in
\end{figure}

Before proceeding to define $p(d)$, we illustrate the idea on a ``shovel" 
in $\R^{D\!=\!3}$ space (Fig.~\ref{fig:illustr1}). The shovel consists of 
2d square blade and 1d handle with uniformly distributed masses 
$M_b \!>\! 0$ and $M_h \!>\! 0$ respectively. If the relevance of space 
points is set by mass they carry, the probability of encountering the handle, the blade 
and the rest of space is $\cP=M_h/(M_b \!+\! M_h)$, $1 \!-\! \cP$ and $0$ 
respectively. Note that UV cutoff $a$ and IR cutoff $L$ are also indicated.

Above we implicitly assumed that $d$ is the usual UV dimension 
($a \!\to\! 0$ at fixed $L$) in which case we have by inspection
$p(d) = \cP\, \delta(d-1) + (1-\cP)\,\delta(d-2)$. But how would this 
$p(d)$ be concluded by a computer that cannot ``see" and only
processes regularized probability vectors $P(a) \!=\! (p_{1},p_{2},\ldots,p_{N(a)})$?
Here $N(a) \!=\! (L/a)^3$, $p_i$ is the probability within elementary cube 
at point $x_i$ of latticized space, and $a \in \{L/k \mid k=2, 3, \ldots \}$.

Anticipating that any number $J$ of discrete dimensions 
$0 \!\le\! d_1 \!<\! d_2 \!<\! \ldots < d_J \!\le\! 3$ with probabilities 
$\cP_j \!>\! 0$ could be present, computer first orders $p_i$ in each 
$P(a)$ so that $p_1 \!\ge p_2 \!\ge \ldots \ge p_{N(a)}$. The rationale is 
that, with decreasing $a$, this increasingly better separates out populations 
related to different $d_j$. Indeed, the typical size of $p$ associated with 
$d_j$ is $\propto\! a^{d_j}$ and so $P(a)$ gradually organizes into $J$ 
sequential blocks starting with $d_1$.  The above ordering in $P$ will 
always be assumed from now on.

To detect possible blocks/dimensions, computer uses variable 
$q \!\in\! [0,1]$ for cumulative probability, and associates with each 
$P(a)$ function $\nu(q,a)$, namely the number of first elements 
in $P(a)$ (space points) whose probabilities add to $q$. Keeping 
track of fractional boundary contributions at each $q$ makes
it a continuous, convex, increasing, piecewise linear function such 
that $\nu(0,a) \!=\!0$ and $\nu(1,a) \!=\! N(a)$. Number of points in 
interval $(q - \epsilon,q \mkern 0.8mu ]$ is $\nu(q,a) - \nu(q-\epsilon,a)$ 
and scales as $a^{-d(q,\epsilon)}$ for $a \!\to\! 0$. When processing 
$P(a)$ for the shovel, computer finds perfect scaling 
$(\ell_h/a) \times \epsilon /\cP$ for $\epsilon \le q \le \cP$, and 
$(\ell_b/a)^2 \times \epsilon / (1-\cP)$ for $\cP + \epsilon < q < 1$. 
It will thus conclude $d(q)$ shown in Fig.~\ref{fig:illustr1} upon 
$\epsilon \to 0$. Value at $q \!=\! 1$ represents the spatial complement 
of the shovel (zero probability). Collecting the probability 
of $d$, namely $p(d) = \int_0^1 dq \,\delta(d-d(q))$ produces the 
inspected~result.

Two points are relevant here. (1) The above approach doesn't 
change if continuous set of dimensions is present. In that case 
the obtained $d(q)$ is not piecewise constant, but rather a piecewise 
continuous non-decreasing function, possibly with constant parts 
identifying discrete dimensions. (2) IR case is fully analogous, but it is useful 
to recall the meaning of IR dimension ($L \!\to\! \infty$, $a$ fixed) which 
is somewhat non-standard. Thus, if both $\ell_h$ and $\ell_b$ are fixed 
as $L \!\to\! \infty$ (usual case) then $p(d) \!=\! \delta(d)$ since populations 
at each $q$ remain constant. However, if e.g. $\ell_b$ is fixed while handle 
responds by $\ell_h \propto L$ (shovel reaches anywhere in space) then 
$p(d) \!=\! (1-\cP) \delta(d) + \cP \delta(d-1)$. 

{\bf 2. The Formalism. $\,$} 
We now define $p(d)$ in IR~setting of Anderson transitions. Such analysis 
pertains to wave functions $\psi \!=\! \psi(r_i)$ on cubic lattice of 
$N(L) \!=\! (L/a)^{D}$ sites $r_i$, with $L$ the IR regulator and $a$ set 
to unity. With $\psi$ we associate the probability vector 
$P\!=\!(p_1, p_2, \ldots, p_{\nrN=\nrN(L)})$, where $p_i = \psi^+\psi(r_i)$, 
the effective number of sites~\cite{Horvath:2018aap, Horvath:2018xgx}
\begin{equation}
      \efNm[\psi]  \,=\, \sum_{i=1}^\nrN \cfu\bigl( \nrN  p_i\bigr)   \quad,\quad
      \cfu(\w)  \; = \;   \min\, \{ \w, 1 \}    \;
     \label{eq:026}         
\end{equation}
and the cumulative count $\nu[q,\psi]$ defined as follows.
Consider cumulative probabilities $(q_0,q_1,\ldots,q_{\nrN})$ with 
$q_0 \!=\! 0$ and $q_j \!=\! \sum_{i=1}^j p(i)$ for $j \!>\!0$. Let $j(q)$, 
$q \!\in\! (0,1)$ be the largest $j$ such that $q_j \!<\! q$. 
Then $\nu[0,\psi] \!=\!0$, $\nu[1,\psi] \!=\! \nrN$~and 
\begin{equation}
     \nu[q,\psi] = j(q) + \frac{q-q_j}{q_{j+1}-q_j} 
     \quad\; , \quad\;  0 < q < 1   \;\;
     \label{eq:036}              
\end{equation}
Recalling the order in $P$, $\nu[q,\psi]$ is increasing and convex.

Consider the Anderson model in orthogonal class~\cite{Anderson:1958a}. 
With $c_{r_i}$ the electron operators, the Hamiltonian is
\begin{equation}
     {\cal H} \,=\, \sum_i \epsilon_{r_i} \, c^\dag_{r_i}  \, c_{r_i} 
     \,+\, \sum_{i,j}  c^\dag_{r_i} \, c_{r_i -e_j} + h.c.
     \label{eq:046}                   
\end{equation}
where $e_j$ ($j \!=\! 1,...,D$) are unit lattice vectors and random 
potentials $\epsilon_{r_i} \!\in\! [-W/2,+W/2]$ are uniformly distributed.
Physics of the model involves averaging over disorder~$\{\epsilon_{r_i}\!\}$. 
For $\efNm$ and $\nu$ of 1-particle eigenstates $\psi$ at energy $E$ we~get
\begin{equation}
    \efNm[\psi] \rightarrow \efNm(E,W,L) 
    \;\;\; , \;\;\;
    \nu[q,\psi]  \rightarrow  \nu(q,E,W,L)
    \;\;  
    \label{eq:056}                  
\end{equation}
Keeping the dependence on $E$, $W$ implicit, $L \!\to\! \infty$ 
behavior defines dimensional characteristics $d_\fir$ and $d(q)$~via
\begin{equation}
   \efNm(L)  \,\propto\,  L^{d_\fir}   
   \;\;\; , \;\;\; 
    \nu(q,L) - \nu(q-\epsilon,L)  \,\propto\,
    L^{d(q,\epsilon)}   \;\;\;
    \label{eq:066}   
\end{equation}
with $d(q) \!=\! \lim_{\epsilon \to 0} d(q,\epsilon)$.
Due to convexity of cumulative counts, $d(q,\epsilon)$ and $d(q)$ are 
non-decreasing. Probability density of finding IR dimension $d$ in 
a state is~then
\begin{equation}
    p(d,\epsilon) =  \int_0^1 d q \,\delta\bigl( d-d(q,\epsilon) \bigr)   \;\;\; , \;\;\;
    p(d) = \lim_{\epsilon \to 0} p(d,\epsilon)    \;\;    
    \label{eq:076}                  
\end{equation}
If $d(q)$ is differentiable at $q$, then $p(d\!=\!d(q)) \!=\! 1/d^{\prime}(q)$.
The range of $d(q)$, equal to support of $p(d)$, specifies IR dimensions 
occurring with non-zero probability in states of interest. 
It is a subset of $[ d_\smin,d_\smax ]$ where
\begin{equation}
     d_\smin  \!=\! \inf\{d \mid p(d) \!>\! 0\}   \;\;,\;\;
     d_\smax  \!=\! \sup\{d \mid p(d) \!>\! 0\}   \;    
     \label{eq:086}                        
\end{equation}

Important feature in the ensuing formalism is that 
\begin{equation}
     d_\smax  \le d_\fir  \le D \;    
     \label{eq:096}                        
\end{equation}
Here the inequalities involving $D$ are obvious and the last
one can be most easily seen in discrete case. Indeed, let
$p(d) \!=\! \sum_{j=1}^J \cP_j \delta(d \!-\! d_j)$ with 
$0 \!\le\! d_1 \!<\! \ldots \!<\! d_J \le D$, $\cP_j \!>\!0$, 
and assume that $d_\fir \!<\! d_J \!=\! d_\smax$. 
Consider $q$ such that  $1 \!-\! \cP_J \!<\! q \!<\! 1$.  
Then $\nu(q,L) \!-\! \nu(q-\epsilon,L) \!=\! 
            \epsilon \,v(q,\epsilon,L) L^{d(q,\epsilon)}$ 
for sufficiently small $\epsilon$, where 
$\lim_{\epsilon \to 0} d(q,\epsilon) \!=\!d_J$ and
$\lim_{\epsilon \to 0}\lim_{L \to \infty} v(q,\epsilon,L) \!=\! v(q) \!>\! 0$.
The size of individual $p \!=\! \epsilon/(\nu(q,L) \!-\! \nu(q-\epsilon,L))$ 
in this population is then $L^{-d(q,\epsilon)}/v(q,L,\epsilon)$. 
Hence, if $d_J \!<\! D$ then $\min \{1,\nrN p\}$ in definition of 
$\efNm$ yields $1$ for sufficient $L$ and $\epsilon$, while if 
$d_J \!=\!D$ it yields $1/v(q)$. In both cases, the contribution of 
this population to $\efNm$ is $\propto \! L^{d_J}$. Hence, 
$d_\fir \ge d_J$ which contradicts the assumption and leads 
to~\eqref{eq:096}.   

\begin{figure}[t]
   \includegraphics[width=0.40\textwidth]{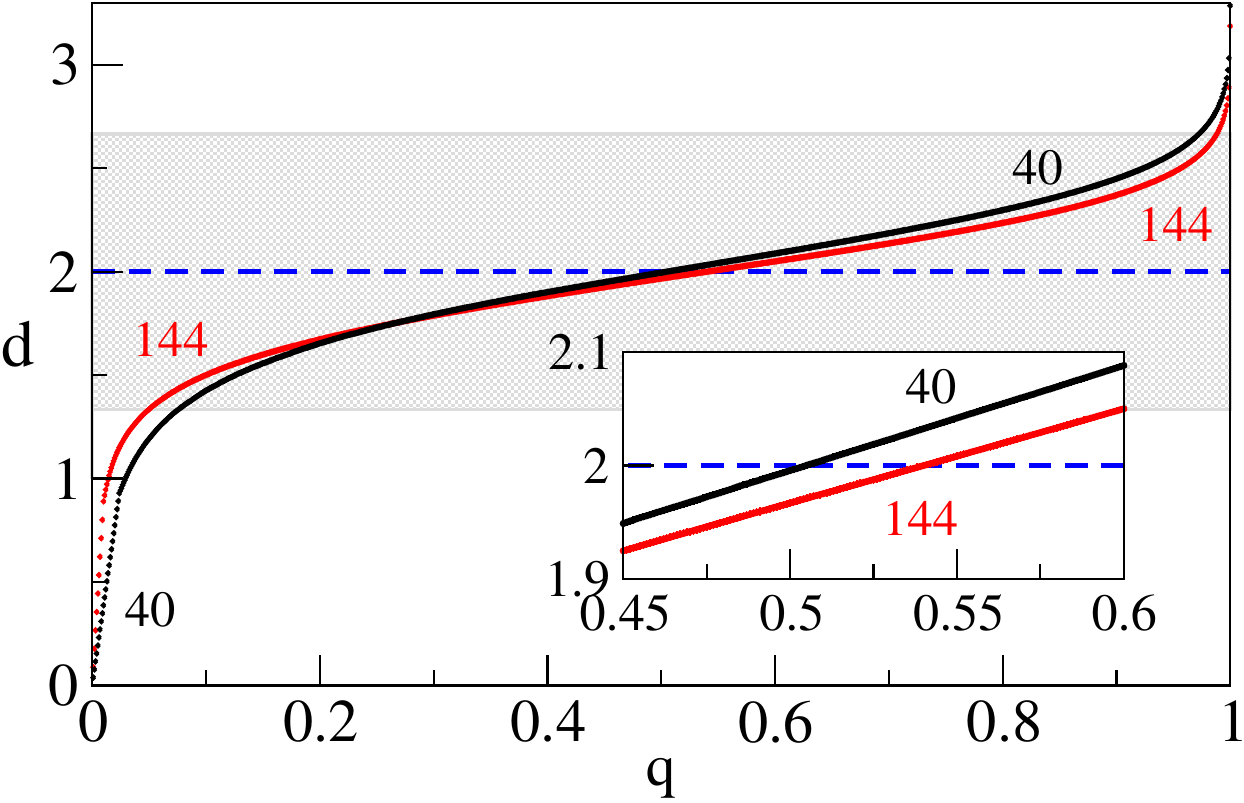}
   \vskip -0.05in   
   \caption{Function $d(q,\epsilon,L)$ at $\epsilon \!=\! 10^{-3}$ 
   for $L \!=\! 40$ and $L\!=\!144$ (largest) systems.
   Shaded region marks the range $d \!\in\! [4/3,8/3]$.}
   \label{fig:dvsq}
   \vskip -0.16in
\end{figure}

{\bf 3. Anderson Criticality. $\,$} 
We now perform the dimensional analysis for critical states of $D\!=\!3$ 
Anderson Hamiltonian~\eqref{eq:026} with periodic boundary conditions 
at critical point $(E_c,W_c) \!=\! (0, 16.543(2))$~\cite{Slevin_2018}. 
Calculation in Ref.~\cite{Horvath:2021zjk} yielded 
$d_\fir \!=\! 2.665(2) \!\approx\! 8/3$. 
For $d(q)$ we follow~\cite{Horvath:2021zjk}, keeping track of dimension 
defined at finite $L$ and extrapolating it directly. In particular,
\begin{equation}
    d(q,\epsilon,L) = \frac{1}{\log s} \,
    \log \frac{\nu(q,L) - \nu(q-\epsilon,L)}{\nu(q,L/s) - \nu(q-\epsilon,L/s)} 
    \;\;\;    
    \label{eq:106}                             
\end{equation}
with fixed $s \!>\!1$, and 
$d(q,\epsilon) \!=\! \lim_{L \to \infty} d(q,\epsilon,L)$. In the analysis we 
set $s\!=\!2$. For 34 sizes in the range $16 \!\le\! L \!\le\! 144$, 
two near-zero eigenmodes were computed at 40k--100k disorder 
samples using the JADAMILU package~\cite{jadamilu_2007}.
We set $\epsilon \!=\!10^{-3}$, thus splitting the interval $q \!\in\! [0,1]$ 
into 1000 bins and evaluating $d(q_b,\epsilon,L)$ at 
$q_b = b \times 10^{-3}$, $b =1,\ldots,1000$. We verified that this is fine 
enough to directly represent $\epsilon \to 0$ limits for our purposes.
 
Given that, we show $d(q,L)$ at $L \!=\! 40$ and $L \!=\! 144$ in 
Fig.~\ref{fig:dvsq}. Important feature of the obtained behavior is the 
flatness in the middle part of $q$, indicating large probabilities for 
dimensions in the corresponding range. Increase of $L$ results in 
flatter $d(q,L)$ and yet sharper range of prominent dimensions. 
Visible linear parts at small $q$ mark regions where finite-size effects 
yield $\nu(q)$ non-convex. Their extent shrinks toward zero with 
growing $L$. Linearity was imposed to keep the behavior regular.

\begin{figure}[t]
   \includegraphics[width=0.35\textwidth]{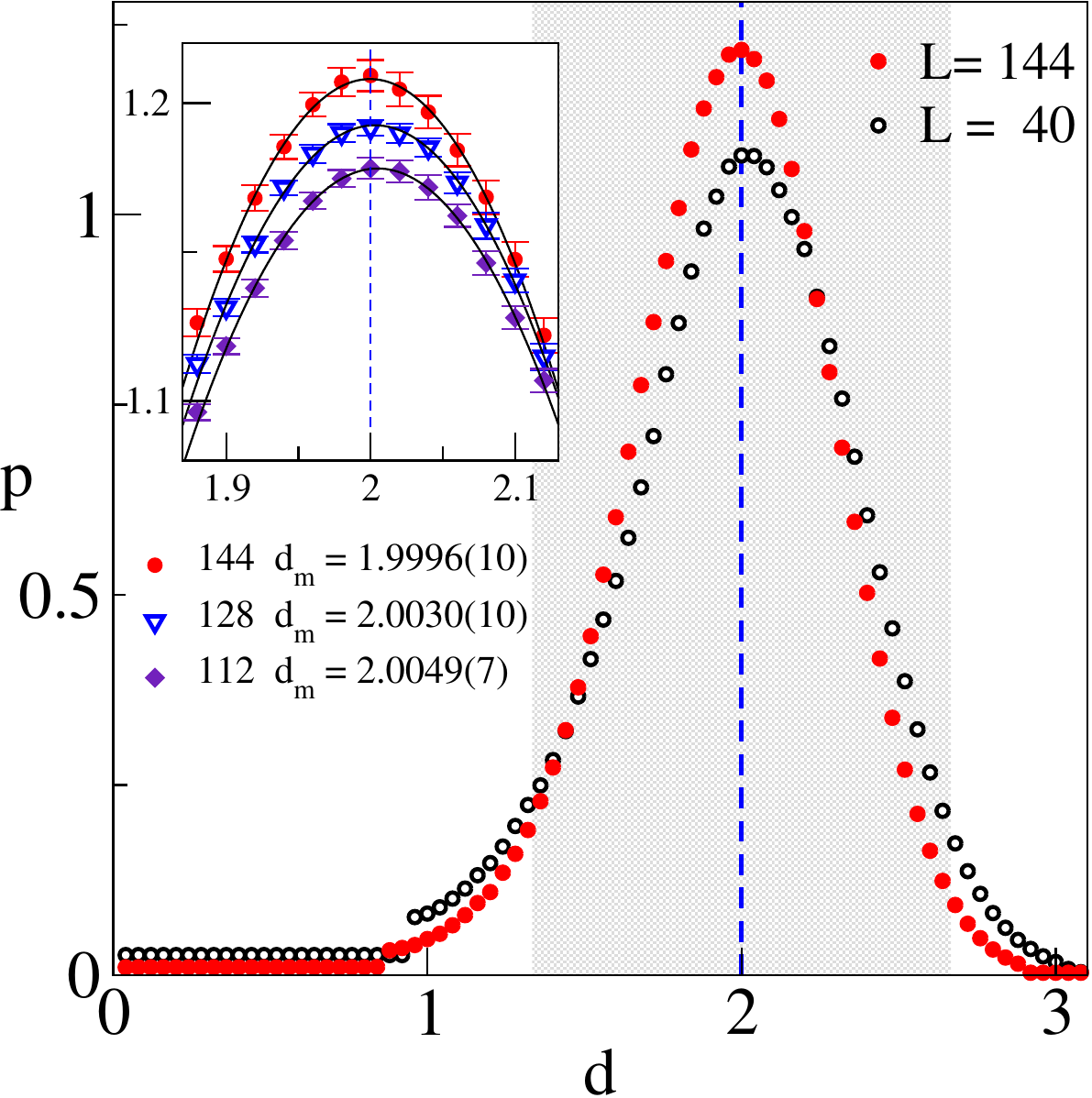}
   \vskip -0.05in   
   \caption{Function $p(d,\epsilon,L)$ at $\epsilon \!=\! 10^{-3}$ for 
   $L \!=\! 40$ and $L \!=\! 144$ (largest) systems. Shaded region 
   marks the range $d \!\in\! [4/3,8/3]$.}
   \label{fig:pvsd}
   \vskip -0.16in
\end{figure}

The corresponding $p(d,L)$ obtained via \eqref{eq:076} are shown 
in Fig.~\ref{fig:pvsd}. We observe sharp peaks of decreasing width,
centered at $d_{\text m} \!\approx\! 2$. The error bars, too small to be 
visible, were obtained via Jackknife procedure with respect to disorder 
samples. Stability of $d_{\text m}$ and its proximity to $2$ is quite remarkable 
as shown in the inset for the largest sizes studied. Quoted values were 
obtained from quadratic fits in the displayed vicinity of the maximum. 
The constant parts at small $d$ correspond to linear 
segments in Fig.~\ref{fig:dvsq}.

\begin{figure}[t]
   \vskip 0.05in
   \includegraphics[width=0.42\textwidth]{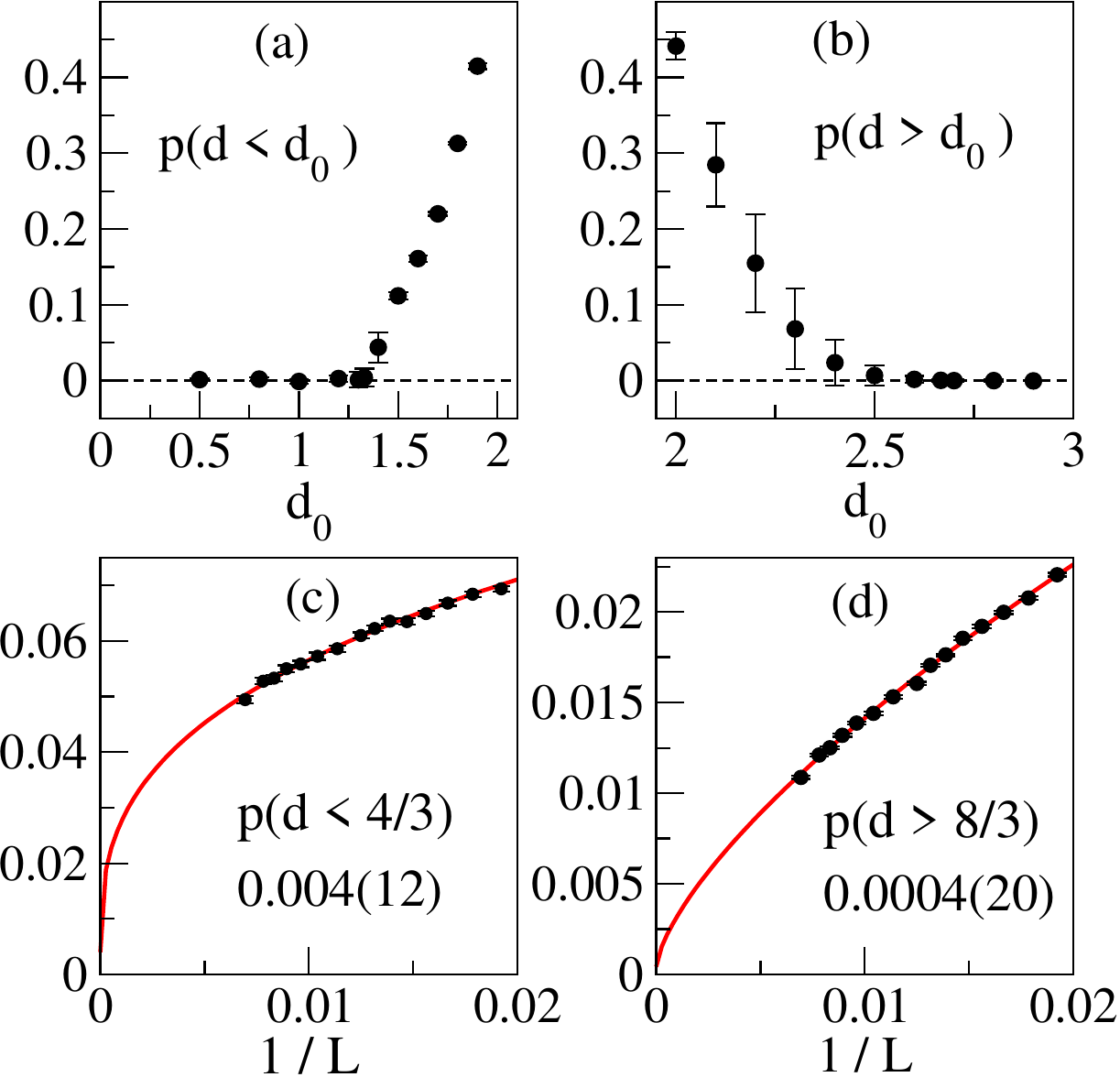}
   \vskip -0.05in   
   \caption{Probabilities $p(d \!<\! d_0)$ and $p( d \!>\! d_0)$ in 
   $L \!\to\! \infty$ limit. Panels (c), (d) show extrapolations for 
   $d_0 \!=\!4/3$ and $d_0 \!=\!8/3$.}
   \label{fig:range}
   \vskip -0.16in
\end{figure}

Among key chracteristics of dimension content $p(d)$ is its support, i.e. 
dimensions that can contribute to physical processes with non-zero 
probability density. The above properties of $p(d,L)$ imply that 
the support in fact spans $[d_\smin, d_\smax]$, and its specification thus 
reduces to finding $d_\smin$ and $d_\smax$. To that effect, we evaluate 
probabilities $p(d \!<\! d_0,L)$ of dimensions smaller than $d_0$, and 
vary $d_0$ upward. For each $d_0$, $p(d \!<\! d_0,L)$ is $L \!\to\! \infty$ 
extrapolated by fitting to a constant with general power correction. 
The result, shown in Fig.~\ref{fig:range} panel (a), features a probability 
threshold turning on near $d_0 \!=\! 1.3$. We take $d_0 \!=\! 4/3$ as 
a reference value: in panel (c) we show its extrapolation leading to a clean 
statistical zero. Analogous procedure based on $p(d \!>\! d_0)$ 
yields results shown in panels (b),(d) with $d_0 \!=\! 8/3$ referencing
the other threshold. 

\begin{figure}[t]
   \vskip 0.05in
   \includegraphics[width=0.35\textwidth]{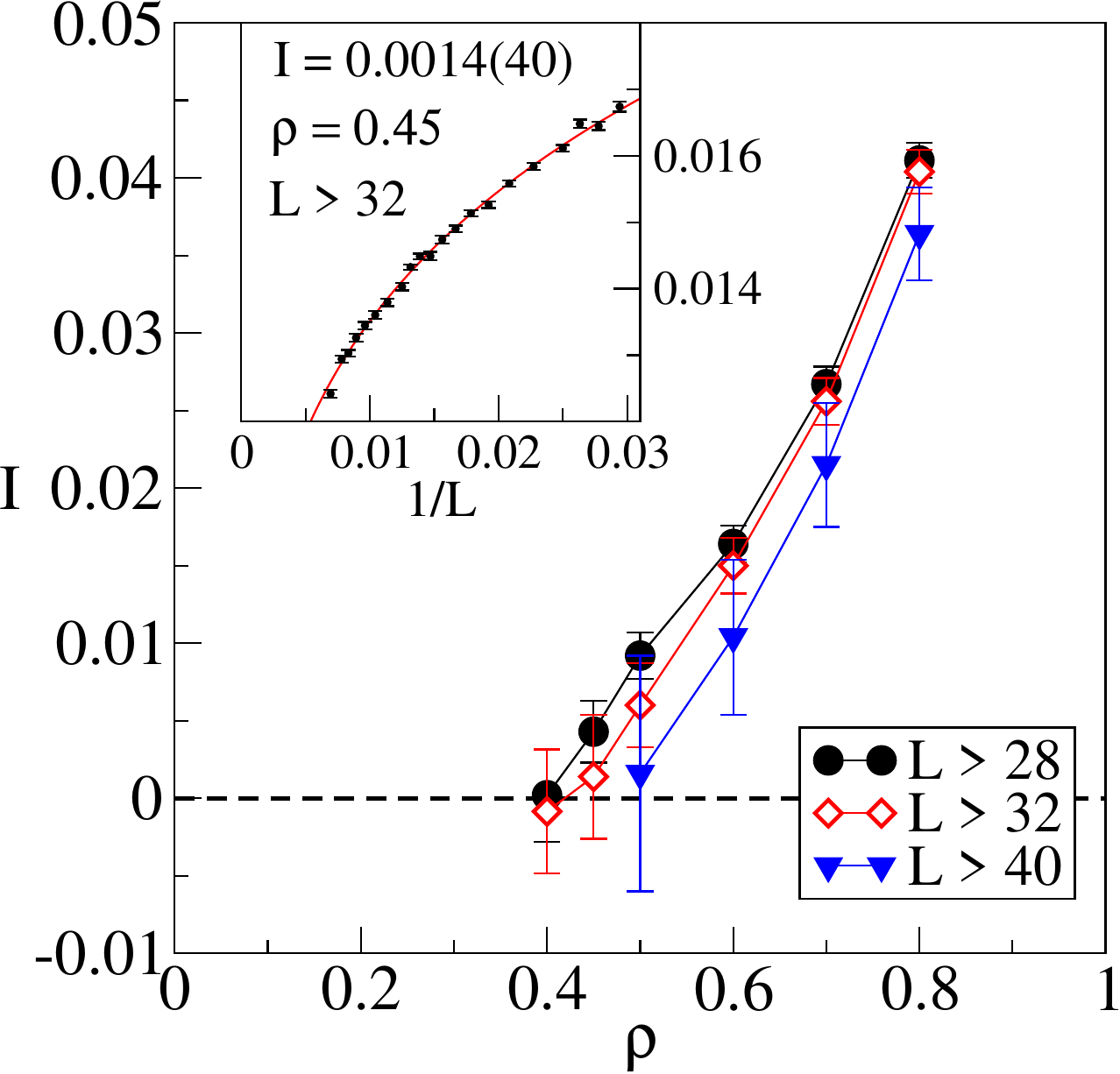}
   \vskip -0.05in   
   \caption{Function $I(\rho, L \!\to\!\infty)$ obtained by fitting in 
   $L$-ranges containing increasingly larger lattices. Inset shows
   example of a fit in the vicinity of $\rho_0$ such that 
   $I(\rho_0) \!\approx\! 0$.}
   \label{fig:discrete}
   \vskip -0.16in
\end{figure}

Given the strong dominance of $d_{\text m}$, the second key question is 
whether $d_{\text m}$ could be a discrete dimension in Anderson critical 
states. This would mean that, in $L \!\to\! \infty$ limit, $d(q,L)$ 
(see Fig.~\ref{fig:dvsq}) develops a strictly constant part in certain 
range of $q$. We will test this possibility for the observed 
$d_{\text m} \!=\! 2$ via the following procedure.  Given a $d(q,L)$, we 
find $q_2(L)$ such that $d(q_2,L)=2$ and calculate
\begin{equation}
     I(\rho,L) = \int_{q_2 - \rho/2}^{q_2 + \rho/2} 
     dq \, \Bigl(\, 2- d(q,L) \,\Bigr)^2
     \label{eq:116}                  
\end{equation} 
which is only zero if $d(q,L) \!=\! 2$ on the interval. For~given $\rho$,
we perform $L \!\to\! \infty$ extrapolation via fit to a constant $I(\rho)$
with general power correction. Fitting data for systems with 
$L \!>\! 28$ leads to results shown in Fig.~\ref{fig:discrete} (circles). 
Notice a steep decay of $I(\rho)$ with decreasing $\rho$, reaching
$I \!\approx\! 0$ at $\rho \!\approx\! 0.4$ with errors becoming large 
below this point. While this could simply indicate a very steep
analytic behavior of $I(\rho)$, further analysis suggests otherwise.
Indeed, restricting fits to larger systems, namely $L \!>\!32$ 
(diamonds) and $L \!>\! 40$ (triangles), results in increasingly steeper 
decay toward zero at yet larger $\rho$. Natural interpretation of these 
tendencies is that $I(\rho) \equiv 0$ for $\rho \!<\! \rho_0 \!\approx\! 0.5$, 
pointing to discrete nature~of~$d_{\text m}$.

\begin{figure}[b]
   \vskip 0.05in
   \includegraphics[width=0.40\textwidth]{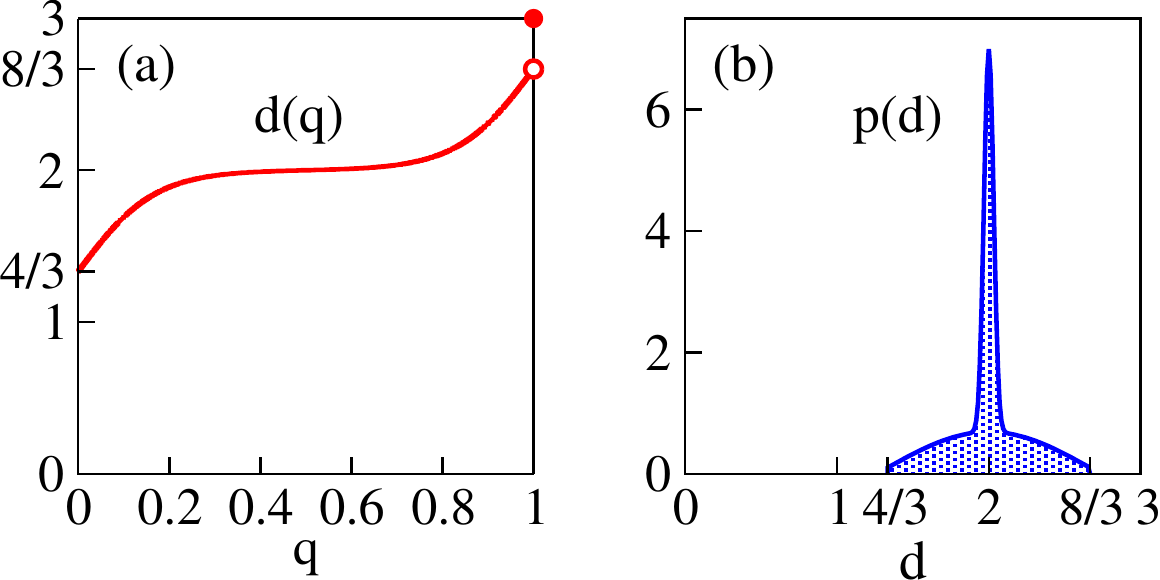}
   \vskip -0.05in   
   \caption{Graphical representation of concluded dimensional content 
   at criticality. Narrow spike in (b) repreents $\delta$-function.}
   \label{fig:concluded}
   \vskip -0.16in
\end{figure}

The synthesis of our results suggests the following form of spatial 
dimension content at Anderson criticality
\begin{equation}
      p(d) = \cP \, \delta(d - d_\text{m})  \,+\, (1-\cP) \, \pi(d)
     \label{eq:126}                   
\end{equation}
where $\pi(d)$ is a continuous probability distribution with support
on interval $[d_\smin,d_\smax]$. The parameters are
\begin{equation} 
     d_{\text m} \!\approx\! 2   
     \;\;\, , \;\;\,     
     d_\smin \!\approx\! 4/3
     \;\;\, , \;\;\,
     d_\smax \!\approx\! 8/3
     \;\;\, , \;\;\,
     \cP \GtrApprox 1/2       
     \;\;
     \label{eq:136}            
\end{equation}
where we estimate the accuracy of $d_m$ at couple 
$\text{\textperthousand}$ and that of $d_\smin$, $d_\smax$ at 
couple $\%$. Graphical representation of this result in terms of
$d(q)$ and $p(d)$ is shown in Fig.~\ref{fig:concluded}. 

{\bf 4. Discussion.} We proposed to characterize probability distributions on 
metric spaces by their measure-based effective dimension ($d_\fuv$ or $d_\fir$) 
\cite{Horvath:2018aap,Horvath:2018xgx, Alexandru:2021pap, Horvath:2022ewv} 
and the associated {\em dimension content} $p(d)$.
The method was applied to the structure of critical states in $D\!=\!3$ Anderson 
transition (O class). Here $p(d)$ identifies dimensions of regions where 
electron can in fact be found, i.e. those relevant to its physics. 
Critical wave functions are subdimensional, multidimensional and our 
new results are summarized by Eqs.~\eqref{eq:126} and \eqref{eq:136}. 
Few comments should be made.

\smallskip
\noindent
{\bf (i)} The picture of Anderson transition as spatial dimension transformation 
\eqref{eq:006} receives key refinements by virtue of $p(d)$. 
Indeed, although critical electron is fully confined to spatial effective support 
$\cS_\star$ of Minkowski dimension 
$d_\fir \!\approx\! 8/3$~\cite{Horvath:2021zjk, Horvath:2022lbj}, its key 
substructure has $d_\text{m} \!\approx\! 2$, and the continuum of lower and 
higher-dimensional features is also present. Geometrically, $\cS_\star$ 
may thus also be viewed as surface-like structure endowed with complex 
lower-dimensional ``hair" and higher-dimensional ``halo".

\vspace{0.02in}

\noindent
{\bf (ii)} Our results suggest that $d_\text{m}$ is a discrete dimension and 
that it may assume an exactly topological value $d_\text{m}\!=\!2$. 
[Mathematical meaning of ``topological" in the context of IR dimension 
would of course need some clarification.] This invokes a possibility that 
quantum mechanics combined with pure disorder can lead 
to emergence of integer dimensions. Apart from understanding
of Anderson transitions, variations on such dynamics could find relevance 
in modeling emergent space in early universe. More detailed 
description of this geometry would be needed.

\vspace{0.02in}

\noindent
{\bf (iii)} Connection between $d_\fir$ and $p(d)$ results from built-in 
additivity which makes them measure-based: in case of $d_\fir$ it is 
additivity of effective counting with respect to combining 
the systems~\cite{Horvath:2018aap,Horvath:2018xgx}, and for
$d(q)$ the familiar additivity of ordinary counting. 
This aspect is key to interpretation of these concepts as spatial 
dimensions. Indeed, it is because Hausdorff measure and Minkowski 
count properly quantify volume that dimensions based on them
became useful and accepted characteristics of space.

\vspace{0.02in}

\noindent
{\bf (iv)} It is natural to ask whether some features of the described 
spatial structure have analogues in the multifractal
approach~\cite{falconer2014fractal, Halsey_Kadanoff_multif:1986}
adopted to IR Anderson setting via moment method~\cite{Evers_2008}.
Here the focus is on the so called dimensional spectrum $f(\alpha)$.
Inner workings of the method give special status to information 
dimension~\cite{Grassberger:1985_1} in a way somewhat similar 
to $d_\text{m}$. It would be interesting to study the possible 
association between the two in detail. (See also debate regarding 
$d_\fir$ in Refs.~\cite{Burmistrov:2022, Horvath:2022com}.)

\vspace{0.02in}

\noindent
{\bf (v)} Our data is consistent with critical wave functions being of proper 
dimension ($d_\fir \!=\! d_\smax$). However, albeit state of the art, 
their statistical power is not sufficient to reach sharper conclusion 
at this point. 

\vspace{0.02in}

\noindent
{\bf (vi)} Our findings acquire another angle in light of recent 
results~\cite{Alexandru:2021pap, Alexandru:2021xoi} in quantum 
chromodynamics (QCD).
The original proposal that Anderson-like mobility edge $\lamA \!>\! 0$ 
appears in QCD Dirac spectrum upon thermal chiral 
transition~\cite{GarciaGarcia:2005vj, GarciaGarcia:2006gr}, worked 
out by Refs.~\cite{Kovacs:2010wx, Giordano:2013taa, Ujfalusi:2015nha}, 
became more structured. Indeed, existence of a new mobility edge 
$\lambda_\fir \!\equiv\! 0$ has been concluded and its simultaneous 
appearance with $\lamA$ at temperature $T_\fir$ was 
conjectured~\cite{Alexandru:2021xoi}. Here $T_\fir$ marks 
the transition to phase featuring IR~scale invariance of glue 
fields~\cite{Alexandru:2019gdm}. Approach to IR criticality 
($\lambda \!\to\! \lambda_\fir^+$) was found to proceed via 
$d_\fir \!\approx\! 2$ Dirac modes~\cite{Alexandru:2021pap}, with 
topological origin of the dimension suspected. Clarifying a possible 
relation of this to $d_\text{m} \!\approx\! 2$ found here may shed new 
light on QCD--Anderson localization connection.  

\vspace{0.02in}

\noindent
{\bf (vii)} The proposed IR/UV guises of multidimensionality formalism 
easily extend to more general situations without metric. Here 
the sequence $\{ \Obj_k \}$ involving collections 
$\Obj_k \!=\! (o_{\scriptscriptstyle{k,1}}, o_{\scriptscriptstyle{k,2}}, 
\ldots, o_{\scriptscriptstyle{k,\nrN_k}})$ 
with increasing number $\nrN_k$ of arbitrary objects comes with 
associated sequence $\{ P_k \}$ of relevance (probability) vectors. 
The role of $d_\fir$/$d_\fuv$ is taken by the effective counting 
dimension $0 \le \Delta \le 1$ defined via scaling
$\efNm[P_k] \propto \nrN_k^{\,\Delta} \;\text{for}\; k \to \infty$ 
\cite{Horvath:2022ewv}. Dimension function $d(q)$ is replaced by 
analogous $\gamma(q)$ and dimension content $p(d)$ by $p(\gamma)$. 
The target ($k \to \infty$) effective collection defined 
by $\{ O_k \}$, $\{ P_k \}$ is then
\begin{equation}
   \begin{aligned} 
   \text{subdimensional if}&  \quad\quad  \Delta < 1 \quad \\
   \text{multidimensional if}&  \quad\quad p(\gamma) \ne \delta(\gamma - \gamma_\smax) \quad \\
   \text{of proper dimension if}&  \quad\quad \Delta = \gamma_\smax 
   \end{aligned}
   \label{eq:146}   
\end{equation}
where $\gamma_\smax = \sup\,\{ \gamma \!\mid\! p(\gamma) > 0 \}$ and
$\gamma_\smax \le \Delta$.
 
\smallskip 
 
\begin{acknowledgments}
   P.M. was supported by Slovak Grant Agency VEGA, Project n. 1/0101/20.
\end{acknowledgments}

\bibliography{my-references}

\begin{thebibliography}{35}%
\makeatletter
\providecommand \@ifxundefined [1]{%
 \@ifx{#1\undefined}
}%
\providecommand \@ifnum [1]{%
 \ifnum #1\expandafter \@firstoftwo
 \else \expandafter \@secondoftwo
 \fi
}%
\providecommand \@ifx [1]{%
 \ifx #1\expandafter \@firstoftwo
 \else \expandafter \@secondoftwo
 \fi
}%
\providecommand \natexlab [1]{#1}%
\providecommand \enquote  [1]{``#1''}%
\providecommand \bibnamefont  [1]{#1}%
\providecommand \bibfnamefont [1]{#1}%
\providecommand \citenamefont [1]{#1}%
\providecommand \href@noop [0]{\@secondoftwo}%
\providecommand \href [0]{\begingroup \@sanitize@url \@href}%
\providecommand \@href[1]{\@@startlink{#1}\@@href}%
\providecommand \@@href[1]{\endgroup#1\@@endlink}%
\providecommand \@sanitize@url [0]{\catcode `\\12\catcode `\$12\catcode
  `\&12\catcode `\#12\catcode `\^12\catcode `\_12\catcode `\%12\relax}%
\providecommand \@@startlink[1]{}%
\providecommand \@@endlink[0]{}%
\providecommand \url  [0]{\begingroup\@sanitize@url \@url }%
\providecommand \@url [1]{\endgroup\@href {#1}{\urlprefix }}%
\providecommand \urlprefix  [0]{URL }%
\providecommand \Eprint [0]{\href }%
\providecommand \doibase [0]{http://dx.doi.org/}%
\providecommand \selectlanguage [0]{\@gobble}%
\providecommand \bibinfo  [0]{\@secondoftwo}%
\providecommand \bibfield  [0]{\@secondoftwo}%
\providecommand \translation [1]{[#1]}%
\providecommand \BibitemOpen [0]{}%
\providecommand \bibitemStop [0]{}%
\providecommand \bibitemNoStop [0]{.\EOS\space}%
\providecommand \EOS [0]{\spacefactor3000\relax}%
\providecommand \BibitemShut  [1]{\csname bibitem#1\endcsname}%
\let\auto@bib@innerbib\@empty
\bibitem [{\citenamefont {Anderson}(1958)}]{Anderson:1958a}%
  \BibitemOpen
  \bibfield  {author} {\bibinfo {author} {\bibfnamefont {P.~W.}\ \bibnamefont
  {Anderson}},\ }\href {\doibase 10.1103/PhysRev.109.1492} {\bibfield
  {journal} {\bibinfo  {journal} {Phys. Rev.}\ }\textbf {\bibinfo {volume}
  {109}},\ \bibinfo {pages} {1492} (\bibinfo {year} {1958})}\BibitemShut
  {NoStop}%
\bibitem [{\citenamefont {Schenk}\ \emph {et~al.}(2008)\citenamefont {Schenk},
  \citenamefont {Bollh{\"o}fer},\ and\ \citenamefont {R{\"o}mer}}]{Romer:2008}%
  \BibitemOpen
  \bibfield  {author} {\bibinfo {author} {\bibfnamefont {O.}~\bibnamefont
  {Schenk}}, \bibinfo {author} {\bibfnamefont {M.}~\bibnamefont
  {Bollh{\"o}fer}}, \ and\ \bibinfo {author} {\bibfnamefont {R.~A.}\
  \bibnamefont {R{\"o}mer}},\ }\href {http://www.jstor.org/stable/20454065}
  {\bibfield  {journal} {\bibinfo  {journal} {SIAM Review}\ }\textbf {\bibinfo
  {volume} {50}},\ \bibinfo {pages} {91} (\bibinfo {year} {2008})}\BibitemShut
  {NoStop}%
\bibitem [{\citenamefont {Abrahams}\ \emph {et~al.}(1979)\citenamefont
  {Abrahams}, \citenamefont {Anderson}, \citenamefont {Licciardello},\ and\
  \citenamefont {Ramakrishnan}}]{Abrahams:1979a}%
  \BibitemOpen
  \bibfield  {author} {\bibinfo {author} {\bibfnamefont {E.}~\bibnamefont
  {Abrahams}}, \bibinfo {author} {\bibfnamefont {P.~W.}\ \bibnamefont
  {Anderson}}, \bibinfo {author} {\bibfnamefont {D.~C.}\ \bibnamefont
  {Licciardello}}, \ and\ \bibinfo {author} {\bibfnamefont {T.~V.}\
  \bibnamefont {Ramakrishnan}},\ }\href {\doibase 10.1103/PhysRevLett.42.673}
  {\bibfield  {journal} {\bibinfo  {journal} {Phys. Rev. Lett.}\ }\textbf
  {\bibinfo {volume} {42}},\ \bibinfo {pages} {673} (\bibinfo {year}
  {1979})}\BibitemShut {NoStop}%
\bibitem [{\citenamefont {Horv\'ath}\ and\ \citenamefont
  {Marko\v{s}}(2022{\natexlab{a}})}]{Horvath:2021zjk}%
  \BibitemOpen
  \bibfield  {author} {\bibinfo {author} {\bibfnamefont {I.}~\bibnamefont
  {Horv\'ath}}\ and\ \bibinfo {author} {\bibfnamefont {P.}~\bibnamefont
  {Marko\v{s}}},\ }\href {\doibase 10.1103/PhysRevLett.129.106601} {\bibfield
  {journal} {\bibinfo  {journal} {Phys. Rev. Lett.}\ }\textbf {\bibinfo
  {volume} {129}},\ \bibinfo {pages} {106601} (\bibinfo {year}
  {2022}{\natexlab{a}})},\ \Eprint {http://arxiv.org/abs/2110.11266}
  {arXiv:2110.11266 [cond-mat.dis-nn]} \BibitemShut {NoStop}%
\bibitem [{\citenamefont {Horv\'ath}\ and\ \citenamefont
  {Mendris}(2020)}]{Horvath:2018aap}%
  \BibitemOpen
  \bibfield  {author} {\bibinfo {author} {\bibfnamefont {I.}~\bibnamefont
  {Horv\'ath}}\ and\ \bibinfo {author} {\bibfnamefont {R.}~\bibnamefont
  {Mendris}},\ }\href {\doibase 10.3390/e22111273} {\bibfield  {journal}
  {\bibinfo  {journal} {Entropy}\ }\textbf {\bibinfo {volume} {22}},\ \bibinfo
  {pages} {1273} (\bibinfo {year} {2020})},\ \Eprint
  {http://arxiv.org/abs/1807.03995} {arXiv:1807.03995 [quant-ph]} \BibitemShut
  {NoStop}%
\bibitem [{\citenamefont {Horv\'ath}(2021)}]{Horvath:2018xgx}%
  \BibitemOpen
  \bibfield  {author} {\bibinfo {author} {\bibfnamefont {I.}~\bibnamefont
  {Horv\'ath}},\ }\href {\doibase 10.3390/quantum3030035} {\bibfield  {journal}
  {\bibinfo  {journal} {Quantum Rep.}\ }\textbf {\bibinfo {volume} {3}},\
  \bibinfo {pages} {534} (\bibinfo {year} {2021})},\ \Eprint
  {http://arxiv.org/abs/1809.07249} {arXiv:1809.07249 [quant-ph]} \BibitemShut
  {NoStop}%
\bibitem [{\citenamefont {Alexandru}\ and\ \citenamefont
  {Horv\'ath}(2021)}]{Alexandru:2021pap}%
  \BibitemOpen
  \bibfield  {author} {\bibinfo {author} {\bibfnamefont {A.}~\bibnamefont
  {Alexandru}}\ and\ \bibinfo {author} {\bibfnamefont {I.}~\bibnamefont
  {Horv\'ath}},\ }\href {\doibase 10.1103/PhysRevLett.127.052303} {\bibfield
  {journal} {\bibinfo  {journal} {Phys. Rev. Lett.}\ }\textbf {\bibinfo
  {volume} {127}},\ \bibinfo {pages} {052303} (\bibinfo {year} {2021})},\
  \Eprint {http://arxiv.org/abs/2103.05607} {arXiv:2103.05607 [hep-lat]}
  \BibitemShut {NoStop}%
\bibitem [{\citenamefont {Horv\'ath}\ \emph {et~al.}(2022)\citenamefont
  {Horv\'ath}, \citenamefont {Marko\v{s}},\ and\ \citenamefont
  {Mendris}}]{Horvath:2022ewv}%
  \BibitemOpen
  \bibfield  {author} {\bibinfo {author} {\bibfnamefont {I.}~\bibnamefont
  {Horv\'ath}}, \bibinfo {author} {\bibfnamefont {P.}~\bibnamefont
  {Marko\v{s}}}, \ and\ \bibinfo {author} {\bibfnamefont {R.}~\bibnamefont
  {Mendris}},\ }\href@noop {} {\  (\bibinfo {year} {2022})},\ \Eprint
  {http://arxiv.org/abs/2205.11520} {arXiv:2205.11520 [hep-lat]} \BibitemShut
  {NoStop}%
\bibitem [{\citenamefont {Horv\'ath}\ and\ \citenamefont
  {Marko\v{s}}(2022{\natexlab{b}})}]{Horvath:2022lbj}%
  \BibitemOpen
  \bibfield  {author} {\bibinfo {author} {\bibfnamefont {I.}~\bibnamefont
  {Horv\'ath}}\ and\ \bibinfo {author} {\bibfnamefont {P.}~\bibnamefont
  {Marko\v{s}}},\ }\href@noop {} {\  (\bibinfo {year} {2022}{\natexlab{b}})},\
  \Eprint {http://arxiv.org/abs/2207.13569} {arXiv:2207.13569
  [cond-mat.dis-nn]} \BibitemShut {NoStop}%
\bibitem [{\citenamefont {Aoki}(1983)}]{Aoki_1983}%
  \BibitemOpen
  \bibfield  {author} {\bibinfo {author} {\bibfnamefont {H.}~\bibnamefont
  {Aoki}},\ }\href {\doibase 10.1088/0022-3719/16/6/007} {\bibfield  {journal}
  {\bibinfo  {journal} {Journal of Physics C: Solid State Physics}\ }\textbf
  {\bibinfo {volume} {16}},\ \bibinfo {pages} {L205} (\bibinfo {year}
  {1983})}\BibitemShut {NoStop}%
\bibitem [{\citenamefont {Soukoulis}\ and\ \citenamefont
  {Economou}(1984)}]{Soukulis:1984a}%
  \BibitemOpen
  \bibfield  {author} {\bibinfo {author} {\bibfnamefont {C.~M.}\ \bibnamefont
  {Soukoulis}}\ and\ \bibinfo {author} {\bibfnamefont {E.~N.}\ \bibnamefont
  {Economou}},\ }\href {\doibase 10.1103/PhysRevLett.52.565} {\bibfield
  {journal} {\bibinfo  {journal} {Phys. Rev. Lett.}\ }\textbf {\bibinfo
  {volume} {52}},\ \bibinfo {pages} {565} (\bibinfo {year} {1984})}\BibitemShut
  {NoStop}%
\bibitem [{\citenamefont {Castellani}\ and\ \citenamefont
  {Peliti}(1986)}]{Castellani:1986}%
  \BibitemOpen
  \bibfield  {author} {\bibinfo {author} {\bibfnamefont {C.}~\bibnamefont
  {Castellani}}\ and\ \bibinfo {author} {\bibfnamefont {L.}~\bibnamefont
  {Peliti}},\ }\href {\doibase 10.1088/0305-4470/19/8/004} {\bibfield
  {journal} {\bibinfo  {journal} {Journal of Physics A: Mathematical and
  General}\ }\textbf {\bibinfo {volume} {19}},\ \bibinfo {pages} {L429}
  (\bibinfo {year} {1986})}\BibitemShut {NoStop}%
\bibitem [{\citenamefont {Evangelou}(1990)}]{Evangelou_1990}%
  \BibitemOpen
  \bibfield  {author} {\bibinfo {author} {\bibfnamefont {S.~N.}\ \bibnamefont
  {Evangelou}},\ }\href {\doibase 10.1088/0305-4470/23/7/006} {\bibfield
  {journal} {\bibinfo  {journal} {Journal of Physics A: Mathematical and
  General}\ }\textbf {\bibinfo {volume} {23}},\ \bibinfo {pages} {L317}
  (\bibinfo {year} {1990})}\BibitemShut {NoStop}%
\bibitem [{\citenamefont {Schreiber}\ and\ \citenamefont
  {Grussbach}(1991)}]{Schreiber:1991a}%
  \BibitemOpen
  \bibfield  {author} {\bibinfo {author} {\bibfnamefont {M.}~\bibnamefont
  {Schreiber}}\ and\ \bibinfo {author} {\bibfnamefont {H.}~\bibnamefont
  {Grussbach}},\ }\href {\doibase 10.1103/PhysRevLett.67.607} {\bibfield
  {journal} {\bibinfo  {journal} {Phys. Rev. Lett.}\ }\textbf {\bibinfo
  {volume} {67}},\ \bibinfo {pages} {607} (\bibinfo {year} {1991})}\BibitemShut
  {NoStop}%
\bibitem [{\citenamefont {Janssen}(1994)}]{Janssen:1994}%
  \BibitemOpen
  \bibfield  {author} {\bibinfo {author} {\bibfnamefont {M.}~\bibnamefont
  {Janssen}},\ }\href {\doibase 10.1142/s021797929400049x} {\bibfield
  {journal} {\bibinfo  {journal} {Int. J. Mod. Phys. B}\ }\textbf {\bibinfo
  {volume} {8}},\ \bibinfo {pages} {943} (\bibinfo {year} {1994})}\BibitemShut
  {NoStop}%
\bibitem [{\citenamefont {Falconer}(2014)}]{falconer2014fractal}%
  \BibitemOpen
  \bibfield  {author} {\bibinfo {author} {\bibfnamefont {K.}~\bibnamefont
  {Falconer}},\ }\href
  {https://www.amazon.com/Fractal-Geometry-Mathematical-Foundations-Applications/dp/111994239X}
  {\emph {\bibinfo {title} {Fractal Geometry: Mathematical Foundations and
  Applications}}},\ \bibinfo {edition} {3rd}\ ed.\ (\bibinfo  {publisher}
  {Wiley},\ \bibinfo {year} {2014})\BibitemShut {NoStop}%
\bibitem [{\citenamefont {Halsey}\ \emph {et~al.}(1986)\citenamefont {Halsey},
  \citenamefont {Jensen}, \citenamefont {Kadanoff}, \citenamefont {Procaccia},\
  and\ \citenamefont {Shraiman}}]{Halsey_Kadanoff_multif:1986}%
  \BibitemOpen
  \bibfield  {author} {\bibinfo {author} {\bibfnamefont {T.~C.}\ \bibnamefont
  {Halsey}}, \bibinfo {author} {\bibfnamefont {M.~H.}\ \bibnamefont {Jensen}},
  \bibinfo {author} {\bibfnamefont {L.~P.}\ \bibnamefont {Kadanoff}}, \bibinfo
  {author} {\bibfnamefont {I.}~\bibnamefont {Procaccia}}, \ and\ \bibinfo
  {author} {\bibfnamefont {B.~I.}\ \bibnamefont {Shraiman}},\ }\href {\doibase
  10.1103/PhysRevA.33.1141} {\bibfield  {journal} {\bibinfo  {journal} {Phys.
  Rev. A}\ }\textbf {\bibinfo {volume} {33}},\ \bibinfo {pages} {1141}
  (\bibinfo {year} {1986})}\BibitemShut {NoStop}%
\bibitem [{\citenamefont {Mildenberger}\ \emph {et~al.}(2002)\citenamefont
  {Mildenberger}, \citenamefont {Evers},\ and\ \citenamefont
  {Mirlin}}]{Mildenberger:2002a}%
  \BibitemOpen
  \bibfield  {author} {\bibinfo {author} {\bibfnamefont {A.}~\bibnamefont
  {Mildenberger}}, \bibinfo {author} {\bibfnamefont {F.}~\bibnamefont {Evers}},
  \ and\ \bibinfo {author} {\bibfnamefont {A.~D.}\ \bibnamefont {Mirlin}},\
  }\href {\doibase 10.1103/physrevb.66.033109} {\bibfield  {journal} {\bibinfo
  {journal} {Physical Review B}\ }\textbf {\bibinfo {volume} {66}},\ \bibinfo
  {pages} {033109} (\bibinfo {year} {2002})}\BibitemShut {NoStop}%
\bibitem [{\citenamefont {Vasquez}\ \emph {et~al.}(2008)\citenamefont
  {Vasquez}, \citenamefont {Rodriguez},\ and\ \citenamefont
  {R{\"o}mer}}]{Vasquez:2008a}%
  \BibitemOpen
  \bibfield  {author} {\bibinfo {author} {\bibfnamefont {L.~J.}\ \bibnamefont
  {Vasquez}}, \bibinfo {author} {\bibfnamefont {A.}~\bibnamefont {Rodriguez}},
  \ and\ \bibinfo {author} {\bibfnamefont {R.~A.}\ \bibnamefont {R{\"o}mer}},\
  }\href {\doibase 10.1103/physrevb.78.195106} {\bibfield  {journal} {\bibinfo
  {journal} {Physical Review B}\ }\textbf {\bibinfo {volume} {78}},\ \bibinfo
  {pages} {195106} (\bibinfo {year} {2008})}\BibitemShut {NoStop}%
\bibitem [{\citenamefont {Rodriguez}\ \emph {et~al.}(2008)\citenamefont
  {Rodriguez}, \citenamefont {Vasquez},\ and\ \citenamefont
  {R{\"o}mer}}]{Rodriguez:2008a}%
  \BibitemOpen
  \bibfield  {author} {\bibinfo {author} {\bibfnamefont {A.}~\bibnamefont
  {Rodriguez}}, \bibinfo {author} {\bibfnamefont {L.~J.}\ \bibnamefont
  {Vasquez}}, \ and\ \bibinfo {author} {\bibfnamefont {R.~A.}\ \bibnamefont
  {R{\"o}mer}},\ }\href {\doibase 10.1103/physrevb.78.195107} {\bibfield
  {journal} {\bibinfo  {journal} {Physical Review B}\ }\textbf {\bibinfo
  {volume} {78}},\ \bibinfo {pages} {195107} (\bibinfo {year}
  {2008})}\BibitemShut {NoStop}%
\bibitem [{\citenamefont {Rodriguez}\ \emph {et~al.}(2011)\citenamefont
  {Rodriguez}, \citenamefont {Vasquez}, \citenamefont {Slevin},\ and\
  \citenamefont {R{\"o}mer}}]{Rodriguez:2011}%
  \BibitemOpen
  \bibfield  {author} {\bibinfo {author} {\bibfnamefont {A.}~\bibnamefont
  {Rodriguez}}, \bibinfo {author} {\bibfnamefont {L.~J.}\ \bibnamefont
  {Vasquez}}, \bibinfo {author} {\bibfnamefont {K.}~\bibnamefont {Slevin}}, \
  and\ \bibinfo {author} {\bibfnamefont {R.~A.}\ \bibnamefont {R{\"o}mer}},\
  }\href {\doibase 10.1103/physrevb.84.134209} {\bibfield  {journal} {\bibinfo
  {journal} {Physical Review B}\ }\textbf {\bibinfo {volume} {84}} (\bibinfo
  {year} {2011}),\ 10.1103/physrevb.84.134209}\BibitemShut {NoStop}%
\bibitem [{\citenamefont {Ujfalusi}\ and\ \citenamefont
  {Varga}(2015)}]{Ujfalusi:2015a}%
  \BibitemOpen
  \bibfield  {author} {\bibinfo {author} {\bibfnamefont {L.}~\bibnamefont
  {Ujfalusi}}\ and\ \bibinfo {author} {\bibfnamefont {I.}~\bibnamefont
  {Varga}},\ }\href {\doibase 10.1103/PhysRevB.91.184206} {\bibfield  {journal}
  {\bibinfo  {journal} {Phys. Rev. B}\ }\textbf {\bibinfo {volume} {91}},\
  \bibinfo {pages} {184206} (\bibinfo {year} {2015})}\BibitemShut {NoStop}%
\bibitem [{\citenamefont {Slevin}\ and\ \citenamefont
  {Ohtsuki}(2018)}]{Slevin_2018}%
  \BibitemOpen
  \bibfield  {author} {\bibinfo {author} {\bibfnamefont {K.}~\bibnamefont
  {Slevin}}\ and\ \bibinfo {author} {\bibfnamefont {T.}~\bibnamefont
  {Ohtsuki}},\ }\href {\doibase 10.7566/jpsj.87.094703} {\bibfield  {journal}
  {\bibinfo  {journal} {Journal of the Physical Society of Japan}\ }\textbf
  {\bibinfo {volume} {87}},\ \bibinfo {pages} {094703} (\bibinfo {year}
  {2018})}\BibitemShut {NoStop}%
\bibitem [{\citenamefont {Bollh\"ofer}\ and\ \citenamefont
  {Notay}(2007)}]{jadamilu_2007}%
  \BibitemOpen
  \bibfield  {author} {\bibinfo {author} {\bibfnamefont {M.}~\bibnamefont
  {Bollh\"ofer}}\ and\ \bibinfo {author} {\bibfnamefont {Y.}~\bibnamefont
  {Notay}},\ }\href@noop {} {\bibfield  {journal} {\bibinfo  {journal} {Comp.
  Phys. Comm.}\ }\textbf {\bibinfo {volume} {177}},\ \bibinfo {pages} {951}
  (\bibinfo {year} {2007})}\BibitemShut {NoStop}%
\bibitem [{\citenamefont {Evers}\ and\ \citenamefont
  {Mirlin}(2008)}]{Evers_2008}%
  \BibitemOpen
  \bibfield  {author} {\bibinfo {author} {\bibfnamefont {F.}~\bibnamefont
  {Evers}}\ and\ \bibinfo {author} {\bibfnamefont {A.~D.}\ \bibnamefont
  {Mirlin}},\ }\href {\doibase 10.1103/revmodphys.80.1355} {\bibfield
  {journal} {\bibinfo  {journal} {Reviews of Modern Physics}\ }\textbf
  {\bibinfo {volume} {80}},\ \bibinfo {pages} {1355} (\bibinfo {year}
  {2008})}\BibitemShut {NoStop}%
\bibitem [{\citenamefont {Grassberger}(1985)}]{Grassberger:1985_1}%
  \BibitemOpen
  \bibfield  {author} {\bibinfo {author} {\bibfnamefont {P.}~\bibnamefont
  {Grassberger}},\ }\href {\doibase
  https://doi.org/10.1016/0375-9601(85)90724-8} {\bibfield  {journal} {\bibinfo
   {journal} {Physics Letters A}\ }\textbf {\bibinfo {volume} {107}},\ \bibinfo
  {pages} {101} (\bibinfo {year} {1985})}\BibitemShut {NoStop}%
\bibitem [{\citenamefont {{Burmistrov}}(2022)}]{Burmistrov:2022}%
  \BibitemOpen
  \bibfield  {author} {\bibinfo {author} {\bibfnamefont {I.~S.}\ \bibnamefont
  {{Burmistrov}}},\ }\href@noop {} {\bibfield  {journal} {\bibinfo  {journal}
  {arXiv e-prints}\ } (\bibinfo {year} {2022})},\ \Eprint
  {http://arxiv.org/abs/2210.10539} {arXiv:2210.10539 [cond-mat.dis-nn]}
  \BibitemShut {NoStop}%
\bibitem [{\citenamefont {{Horv{\'a}th}}\ and\ \citenamefont
  {{Marko{\v{s}}}}(2022)}]{Horvath:2022com}%
  \BibitemOpen
  \bibfield  {author} {\bibinfo {author} {\bibfnamefont {I.}~\bibnamefont
  {{Horv{\'a}th}}}\ and\ \bibinfo {author} {\bibfnamefont {P.}~\bibnamefont
  {{Marko{\v{s}}}}},\ }\href@noop {} {\bibfield  {journal} {\bibinfo  {journal}
  {arXiv e-prints}\ } (\bibinfo {year} {2022})},\ \Eprint
  {http://arxiv.org/abs/2212.02912} {arXiv:2212.02912 [cond-mat.dis-nn]}
  \BibitemShut {NoStop}%
\bibitem [{\citenamefont {Alexandru}\ and\ \citenamefont
  {Horv\'ath}(2022)}]{Alexandru:2021xoi}%
  \BibitemOpen
  \bibfield  {author} {\bibinfo {author} {\bibfnamefont {A.}~\bibnamefont
  {Alexandru}}\ and\ \bibinfo {author} {\bibfnamefont {I.}~\bibnamefont
  {Horv\'ath}},\ }\href {\doibase 10.1016/j.physletb.2022.137370} {\bibfield
  {journal} {\bibinfo  {journal} {Phys. Lett. B}\ }\textbf {\bibinfo {volume}
  {833}},\ \bibinfo {pages} {137370} (\bibinfo {year} {2022})},\ \Eprint
  {http://arxiv.org/abs/2110.04833} {arXiv:2110.04833 [hep-lat]} \BibitemShut
  {NoStop}%
\bibitem [{\citenamefont {Garcia-Garcia}\ and\ \citenamefont
  {Osborn}(2006)}]{GarciaGarcia:2005vj}%
  \BibitemOpen
  \bibfield  {author} {\bibinfo {author} {\bibfnamefont {A.~M.}\ \bibnamefont
  {Garcia-Garcia}}\ and\ \bibinfo {author} {\bibfnamefont {J.~C.}\ \bibnamefont
  {Osborn}},\ }\href {\doibase 10.1016/j.nuclphysa.2006.02.011} {\bibfield
  {journal} {\bibinfo  {journal} {Nucl. Phys. A}\ }\textbf {\bibinfo {volume}
  {770}},\ \bibinfo {pages} {141} (\bibinfo {year} {2006})},\ \Eprint
  {http://arxiv.org/abs/hep-lat/0512025} {arXiv:hep-lat/0512025} \BibitemShut
  {NoStop}%
\bibitem [{\citenamefont {Garcia-Garcia}\ and\ \citenamefont
  {Osborn}(2007)}]{GarciaGarcia:2006gr}%
  \BibitemOpen
  \bibfield  {author} {\bibinfo {author} {\bibfnamefont {A.~M.}\ \bibnamefont
  {Garcia-Garcia}}\ and\ \bibinfo {author} {\bibfnamefont {J.~C.}\ \bibnamefont
  {Osborn}},\ }\href {\doibase 10.1103/PhysRevD.75.034503} {\bibfield
  {journal} {\bibinfo  {journal} {Phys. Rev. D}\ }\textbf {\bibinfo {volume}
  {75}},\ \bibinfo {pages} {034503} (\bibinfo {year} {2007})},\ \Eprint
  {http://arxiv.org/abs/hep-lat/0611019} {arXiv:hep-lat/0611019} \BibitemShut
  {NoStop}%
\bibitem [{\citenamefont {Kovacs}\ and\ \citenamefont
  {Pittler}(2010)}]{Kovacs:2010wx}%
  \BibitemOpen
  \bibfield  {author} {\bibinfo {author} {\bibfnamefont {T.~G.}\ \bibnamefont
  {Kovacs}}\ and\ \bibinfo {author} {\bibfnamefont {F.}~\bibnamefont
  {Pittler}},\ }\href {\doibase 10.1103/PhysRevLett.105.192001} {\bibfield
  {journal} {\bibinfo  {journal} {Phys. Rev. Lett.}\ }\textbf {\bibinfo
  {volume} {105}},\ \bibinfo {pages} {192001} (\bibinfo {year} {2010})},\
  \Eprint {http://arxiv.org/abs/1006.1205} {arXiv:1006.1205 [hep-lat]}
  \BibitemShut {NoStop}%
\bibitem [{\citenamefont {Giordano}\ \emph {et~al.}(2014)\citenamefont
  {Giordano}, \citenamefont {Kovacs},\ and\ \citenamefont
  {Pittler}}]{Giordano:2013taa}%
  \BibitemOpen
  \bibfield  {author} {\bibinfo {author} {\bibfnamefont {M.}~\bibnamefont
  {Giordano}}, \bibinfo {author} {\bibfnamefont {T.~G.}\ \bibnamefont
  {Kovacs}}, \ and\ \bibinfo {author} {\bibfnamefont {F.}~\bibnamefont
  {Pittler}},\ }\href {\doibase 10.1103/PhysRevLett.112.102002} {\bibfield
  {journal} {\bibinfo  {journal} {Phys. Rev. Lett.}\ }\textbf {\bibinfo
  {volume} {112}},\ \bibinfo {pages} {102002} (\bibinfo {year} {2014})},\
  \Eprint {http://arxiv.org/abs/1312.1179} {arXiv:1312.1179 [hep-lat]}
  \BibitemShut {NoStop}%
\bibitem [{\citenamefont {Ujfalusi}\ \emph {et~al.}(2015)\citenamefont
  {Ujfalusi}, \citenamefont {Giordano}, \citenamefont {Pittler}, \citenamefont
  {Kov\'acs},\ and\ \citenamefont {Varga}}]{Ujfalusi:2015nha}%
  \BibitemOpen
  \bibfield  {author} {\bibinfo {author} {\bibfnamefont {L.}~\bibnamefont
  {Ujfalusi}}, \bibinfo {author} {\bibfnamefont {M.}~\bibnamefont {Giordano}},
  \bibinfo {author} {\bibfnamefont {F.}~\bibnamefont {Pittler}}, \bibinfo
  {author} {\bibfnamefont {T.~G.}\ \bibnamefont {Kov\'acs}}, \ and\ \bibinfo
  {author} {\bibfnamefont {I.}~\bibnamefont {Varga}},\ }\href {\doibase
  10.1103/PhysRevD.92.094513} {\bibfield  {journal} {\bibinfo  {journal} {Phys.
  Rev. D}\ }\textbf {\bibinfo {volume} {92}},\ \bibinfo {pages} {094513}
  (\bibinfo {year} {2015})},\ \Eprint {http://arxiv.org/abs/1507.02162}
  {arXiv:1507.02162 [cond-mat.dis-nn]} \BibitemShut {NoStop}%
\bibitem [{\citenamefont {Alexandru}\ and\ \citenamefont
  {Horv\'ath}(2019)}]{Alexandru:2019gdm}%
  \BibitemOpen
  \bibfield  {author} {\bibinfo {author} {\bibfnamefont {A.}~\bibnamefont
  {Alexandru}}\ and\ \bibinfo {author} {\bibfnamefont {I.}~\bibnamefont
  {Horv\'ath}},\ }\href {\doibase 10.1103/PhysRevD.100.094507} {\bibfield
  {journal} {\bibinfo  {journal} {Phys. Rev. D}\ }\textbf {\bibinfo {volume}
  {100}},\ \bibinfo {pages} {094507} (\bibinfo {year} {2019})},\ \Eprint
  {http://arxiv.org/abs/1906.08047} {arXiv:1906.08047 [hep-lat]} \BibitemShut
  {NoStop}%
\end{thebibliography}%

\end{document}